# ANALYSIS OF ONE ASSUMPTION OF THE NAVIER-STOKES EQUATIONS

**V.A. Budarin**

*National Polytechnical University, Odesa, Ukraine*

This article analyses the assumptions regarding the influence of pressure forces during the calculation of the motion of a Newtonian fluid. The purpose of the analysis is to determine the reasonableness of the assumptions and their impact on the results of the analytical calculation. The connections between equations, causes of discrepancies in exact solutions of the Navier-Stokes equations at low Reynolds numbers and the emergence of unstable solutions using computer programs are also addressed. The necessity to complement the well-known equations of motion in mechanical stress requires other equations are substantive. It is shown that there are three methods of solving such a problem and the requirements for the unknown equations are described.

## 1. Introduction

В основе теоретического описания многих процессов течения жидкостей, взаимодействия текучей среды и твердого тела во внешней и внутренней задаче гидродинамики, а также при расчете многих других процессов, лежит система уравнений движения вязкой жидкости в напряжениях (Навье) (1) [1]. На протяжении большого периода времени удалось получить малое количество точных решений простых частных задач, которые имеют одно общее свойство - решения справедливы при весьма малых числах Рейнольдса. Например, одна из задач о вязком течении на плоскости, возникающем при контакте с ней вихревой нити справедлива при числе Re < 5,5. Для других задач эти числа Рейнольдса значительно меньше [2]. В тепловой энергетике и во многих других областях такие числа Рейнольдса отсутствуют, что привело к разработке других методов расчета практически важных задач, среди которых выделим путь, основанный на расширении возможностей точных решений. Например, в работе [3] рассматривается возможность распространения точного решения о течении вязкой затопленной струи Ландау на гидродинамические и акустические течения. Данный путь не получил широкого распространения в инженерной практике, однако является полезным, например, при проведении сравнительных и тестовых расчетов.

   Другим направлением использования системы уравнений Навье-Стокса, которое имеет четыре неизвестных, является ее замыкание с помощью полуэмпирических уравнений теории турбулентности с последующим ее численным решением. Такое направление оказалось весьма плодотворным и для его реализации разработаны специализированные математические пакеты Fluent, Esi_CFDrc, Flow Vision и др. Получаемые результаты характеризуются хорошей наглядностью, но одновременно являются неустойчивыми и требуют сравнения результатов, полученных разными методами, в том числе и с помощью эксперимента. Тем не менее, электронные базы данных решенных задач постоянно пополняются и используются проектировщиками.

Еще одним направлением получения уравнений движения является составление расчетной схемы частной задачи и применение к ней частного случая уравнения Навье-Стокса или уравнения равновесия, следующего из правил теоретической механики. Примером такой задачи является течение Пуазейля. Характерной особенностью получаемого этим путем результата является возможность применения решения для всех чисел Рейнольдса, вплоть до возникновения турбулентности. Это означает, что малые числа Рейнольдса в классических решениях не характеризуют физические свойства процесса, а отражают свойство (недостаток) математической модели. Данный путь получения уравнений движения требует хорошего понимания частного процесса для составления корректной расчетной схемы. Кроме того, существует вероятность существования скрытых допущений, что требует проведения эксперимента или сравнительных расчетов.

Анализ рассмотренных выше проблем, показывает, что ключевой задачей совершенствования методов расчета течений различных классов, является выяснение причин возникновения ограничений по числу Рейнольдса для всех частных решений уравнения Навье-Стокса.

Одним из возможных ответов на этот вопрос является анализ одного из допущений, которое касается уравнения для вычисления среднего давления $p$. Принятое линейное уравнение для вычисления $p$ является допущением, которое оправдывается практикой [1].

В настоящей статье обсуждается данное допущение и оценивается его влияние на точные решения.

## 2. Comparision of equations

*2.1. Уравнение Навье*

Уравнение Навье-Стокса предназначено для описания течения наиболее распространенного частного вида текучей среды – вязкой ньютоновской жидкости. Основой для вывода этого уравнения послужило уравнение движения в напряжениях (Навье), которое в координатной форме имеет вид

$$X + \frac{1}{\rho}\left(\frac{\partial p_{xx}}{\partial x} + \frac{\partial \tau_{yx}}{\partial y} + \frac{\partial \tau_{zx}}{\partial z}\right) = \frac{du_x}{dt}$$

$$Y + \frac{1}{\rho}\left(\frac{\partial \tau_{xy}}{\partial x} + \frac{\partial p_{yy}}{\partial y} + \frac{\partial \tau_{zy}}{\partial z}\right) = \frac{du_y}{dt} \qquad (1)$$

$$Z + \frac{1}{\rho}\left(\frac{\partial \tau_{xz}}{\partial x} + \frac{\partial \tau_{yz}}{\partial y} + \frac{\partial p_{zz}}{\partial z}\right) = \frac{du_z}{dt}$$

где  $X, Y, Z$ – удельные массовые силы, $\frac{\text{м}}{\text{с}^2}$ ; $p_{xx}, p_{yy}, p_{zz}$ – проекции нормального напряжения на оси координат, Па; $\rho$ - плотность, $\frac{\text{кг}}{\text{м}^3}$ ; $\tau_{ij}$ – касательные напряжения, Па; $\frac{du_x}{dt} = \frac{\partial u_x}{\partial t} + \frac{\partial^2 u_x}{\partial x^2} + \frac{\partial^2 u_y}{\partial x^2} + \frac{\partial^2 u_z}{\partial x^2}$  - полное ускорение частицы вдоль оси х; $u_x, u_y, u_z$ – проекции скорости на оси координат.

Уравнение (1) имеет девять неизвестных и требует для своего замыкания еще шесть уравнений.

*2.2. Уравнение Навье-Стокса*

Уравнение Навье-Стокса было получено с целью сокращения количества неизвестных и было более пригодно для практического использования. В результате была получена следующая система уравнений для несжимаемой ньютоновской жидкости [1, 4].

$$X - \frac{1}{\rho}\frac{\partial p}{\partial x} + \nu \cdot \nabla^2 u_x = \frac{du_x}{dt}$$

$$Y - \frac{1}{\rho}\frac{\partial p}{\partial y} + \nu \cdot \nabla^2 u_y = \frac{du_y}{dt} \qquad (2)$$

$$Z - \frac{1}{\rho}\frac{\partial p}{\partial z} + \nu \cdot \nabla^2 u_z = \frac{du_z}{dt}$$

где $p_x, p_y, p_z$ – проекции давления на оси координат, $\nu$ - кинематическая вязкость, $\frac{\text{м}^2}{\text{с}}$.

Эта система уравнений имеет четыре неизвестных и также незамкнута, как и система (1). Уравнения (1) и (2) являются нелинейным в связи с присутствием в правой части трех конвективных ускорений.

### 3. Brife analysis of the system (1) and (2)

**1**. Отличие систем в части нормальных сил состоит в разных результатах решения уравнений. Решив систему (1) мы получим компоненты нормального напряжения $p_{xx}$, $p_{yy}$, $p_{zz}$ , т.е. конечный результат расчета. Решив систему (2) мы получим промежуточный результат, который требует пересчета величины   $p$  в компоненты давления $p_x = -p_{xx}$, $p_y = -p_{yy}$, $p_z = -p_{zz}$.   Этот пересчет выполняется с помощью следующего линейного уравнения [1].

$$p = \frac{p_x + p_y + p_z}{3} \qquad (3)$$

Таким образом, уравнение (3) предполагает, что расчетная величина $p$ (давление) является средним арифметическим от искомых компонент давления. В итоге имеет место противоречие между нелинейным  характером уравнения (2) и предположением о линейном  законе осреднения результатов расчета. В общем случае такое допущение является некорректным и не может привести к положительным результатам расчета процесса течения. Кроме того, зависимость (3) с точность до константы выражает теорему суперпозиции решений, которая верна только для линейных уравнений.

Задача осреднения результатов расчетов возникает и в других областях науки. Например, в технической термодинамике и теории теплообмена существует много задач, в которых результаты решения нелинейных уравнений осредняются по нелинейным формулам [5,6].

Однако для некоторых частных задач такое допущение (уравнение 3) можно считать справедливым, но при использовании дополнительного допущения. Необходимо принять, что линейный закон осреднения (3) верен для нелинейной функции. Такое допущение выполняется для задач в которых можно пренебречь конвективными ускорениями или при сокращении интервала осреднения. Геометрическая интерпретация последнего действия – замена кривой на отрезок прямой.

**2**.Уравнение (3) содержит семантическую неточность, т.к. является приближенным. Более правильным является использование в уравнении (3) знака приближенного равенства. Отмеченное противоречие приводит к изменению статуса системы Навье-Стокса с точного на приближенное.

**3**.Требование малости интервала осреднения для корректности уравнений (3) означает, что система (2) должна стать линейной, т.е. в правой части должна присутствовать только частная производная по времени, а конвективные ускорения должны стремиться к нулю. Это возможно, если компоненты скорости в выражениях для конвективных ускорений являются малыми, что соответствует малым числам Рейнольдса.

В настоящее время известные точные решения этой системы фактически подтверждают указанный вывод, т.к. они справедливы при Re ≈ $10^{-3}$… 5,5 [2,7].

### 4.Motion equations and its particular cases

В настоящее время общепризнанными являются три модели текучей среды: идеальная, невязкая и, наиболее общая, вязкая жидкость. Все эти модели взаимосвязаны между собой системой условий и любые дополнительные уравнения, претендующие на роль замыкающих, должны учитывать эти взаимосвязи. На рисунке 1 показана взаимосвязь условий и математических моделей текучей среды, которые следуют из общего уравнения движения.

В соответствии со схемой рис.1 из уравнения (1) можно получить частные случаи уравнений движения для трех моделей текучей среды. При этом уравнение Эйлера можно получить двумя путями: из уравнения Навье-Стокса (по условию *ν = 0*) и с помощью уравнения (1) по двум условиям. Последний путь является более корректным т.к. он приводит к заключению, что уравнение Эйлера является точным, а не приближенным, как следует из первого пути.

Модель невязкой жидкости должна считаться точной, так как никак не связана с уравнением Навье-Стокса, содержит шесть неизвестных и является незамкнутой. Тем не менее, данная модель может использоваться для решения некоторых частных задач. Пример использования этой модели текучей среды для расчета разрывного течения вращающейся жидкости с двумя поверхностями давления приведен в работе [8].

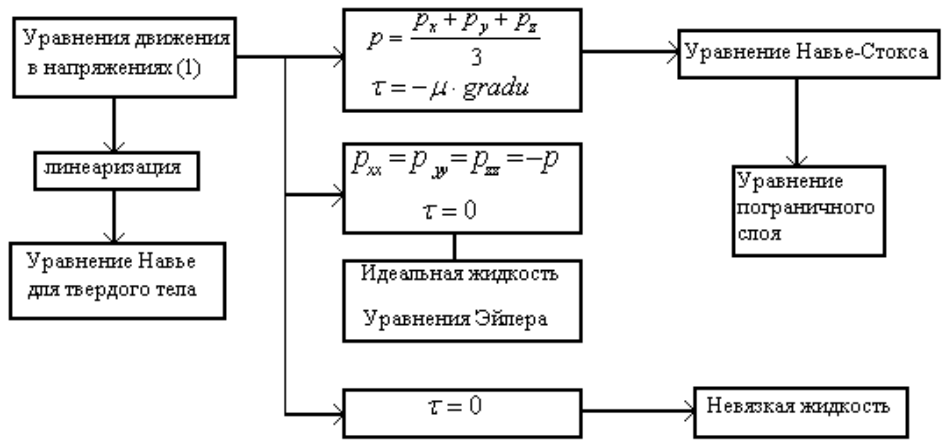

**Рис. 1.** Схема взаимосвязей условий и уравнений движения для различных моделей текучих сред.

### 5. Analysis of the system equations for the flow

**1**. Современные пакеты прикладных программ для расчета процессов течения жидкостей и газов используют уравнение Навье-Стокса совместно с различными приближенными (полуэмпирическими) уравнениями теории турбулентности, которые позволяют замкнуть уравнение движения. В итоге мы получаем систему, все уравнения в которой приближенные. Решение такой системы также должно считаться приближенным, что может проявиться в непрогнозируемых (нустойчивых) результатах.

**2**. Из анализа допущений следует, что существенно улучшить возможности системы уравнений п.1 не представляется возможным. Более рациональным следует считать путь составления замкнутой системы уравнений на основе уравнений (1), дополнив ее другими уравнениями, по аналогии с используемыми в другой области механики сплошной среды – теории упругости [9].

**3**. Взаимосвязь различных уравнений такой системы можно иллюстрировать схемой рисунка 2. Поле напряжений содержит три уравнения, т.к. вектор силы можно разложить на три составляющих. Поле скоростей деформаций содержит шесть уравнений, т.к. разкладываются главные векторы линейных и угловых деформаций. В итоге получаем девять уравнений, образующих замкнутую систему. Из схемы также следует, что некоторые частные задачи можно решать с использованием уравнений для любого поля с последующим пересчетом результатов в параметры второго поля, пользуясь уравнениями связи.

**4**. Известная система из шести уравнений, приведенная в теории деформационного движения, выведена на основе правил векторной алгебры и использует понятие скорости течения. При использовании этой системы уравнений для аналитического решения частных задач необходимо вводить ограничения на линейные и угловые скорости течения. Таким образом, такая система уравнений описывает поведение неголономной механической системы. Практические результаты использования этой системы уравнений отсутствуют, что косвенно указывает на неправомерность ее использования для точного решения частных задач.

**5**. В настоящее время задачи течения жидкости решаются с использованием преимущественно уравнений для поля давлений, что объясняется отсутствием всех необходимых уравнений для описания течени

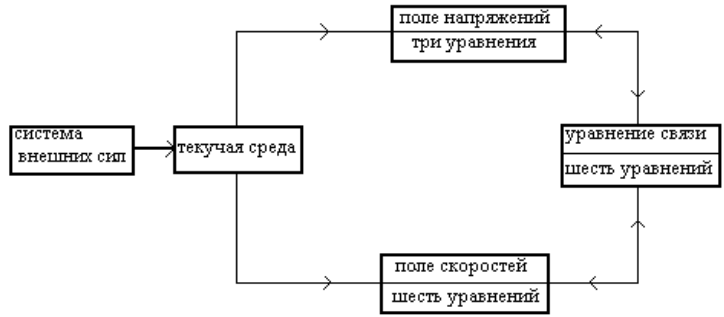

**Рис. 2.** Физические поля и максимальное число уравнений для их описания

## 6. Conclusions

**1.** В результате анализа схемы вывода системы уравнений Навье-Стокса найдено противоречие, которое требует изменения статуса системы с точного на приближенное.
**2**.Любая задача может быть решена с использованием одного из трех методов в основе которых лежит система из двух уравнений:
1.Основная система включает девять уравнений, описывающих силовое поле и поле скоростей.
2.Дополнительная система включает уравнения силового поля и уравнения связи.
3.Дополнительная система включает уравнения поля скоростей и уравнения связи.
**3**.Для составления замкнутых систем уравнений необходимо написать и разработать методы использования еще двух систем уравнений – уравнения связи и общие уравнения для поля скорости. Последнюю систему необходимо вывести в предположении, что движущаяся жидкость представляет собой голономную механическую систему.